\documentclass[12pt]{article}
\usepackage{amsfonts}

\def\ltwid{\mathrel{\raise.3ex\hbox{$<$\kern-.75em\lower1ex\hbox{$\sim$}}}}
\def \be{\begin{equation}}
\def \ee{\end{equation}}
\def \bea{\begin{eqnarray}}
\def \eea{\end{eqnarray}}

\begin{document}

\begin{titlepage}

\begin{flushright}
UFIFT-QG-07-05
\end{flushright}

\vspace{2cm}

\begin{center}
{\bf Scalar Field Equations from Quantum Gravity during Inflation}
\end{center}

\vspace{.5cm}

\begin{center}
E. O. Kahya$^{\dagger}$ and R. P. Woodard$^{\ddagger}$
\end{center}

\vspace{.5cm}

\begin{center}
\it{Department of Physics \\
University of Florida \\
Gainesville, FL 32611}
\end{center}

\vspace{1cm}

\begin{center}
ABSTRACT
\end{center}
We exploit a previous computation of the self-mass-squared from
quantum gravity to include quantum corrections to the scalar evolution
equation. The plane wave mode functions are shown to receive no significant
one loop corrections at late times. This result probably applies as well to
the inflaton of scalar-driven inflation. If so, there is no significant
correction to the $\varphi \varphi$ correlator that plays a crucial role
in computations of the power spectrum.

\vspace{2cm}

\begin{flushleft}
\end{flushleft}

\vspace{.4cm}
\begin{flushleft}
$^{\dagger}$ e-mail: emre@phys.ufl.edu \\
$^{\ddagger}$ e-mail: woodard@phys.ufl.edu
\end{flushleft}
\end{titlepage}

\section{Introduction}

Massless, minimally coupled scalars and gravitons are unique in
possessing zero mass without classical conformal invariance. Simple
arguments based on the energy-time uncertainty principle suggest
that quantum effects from these particles should be vastly enhanced
during primordial inflation \cite{RPW1}. The lowest order effect is
that inflation rips long wavelength virtual quanta of these types
out of the vacuum. That is the origin of the observed scalar
perturbations \cite{MC} and of the potentially observable tensor
perturbations \cite{AAS1}.

Higher order effects derive from the interactions of the sea of
long wavelength virtual scalars and gravitons, both with
themselves and with other particles. Loop corrections to the
primordial power spectra are enhanced by logarithms of the ratio
of the scale factor to its value at first horizon crossing
\cite{SW,BP}. For any mode whose spatial variation we can resolve
today these secular enhancements cannot overcome the minuscule
loop counting parameter of $G H^2 \ltwid 10^{-12}$. Exceptions
have been suggested \cite{DS,VMS,Sloth}. However, much larger loop
corrections (to other things) can be obtained either from
interactions with larger loop counting parameters, or by studying
things we would perceive as spatially constant such as the vacuum
energy or particle masses.

Explicit computations have been made on de Sitter background in five
different models which involve either scalars or gravitons:
\begin{enumerate}
\item{For a massless, minimally coupled scalar with a quartic
self-interaction --- the two loop expectation value of the stress
tensor \cite{OW} and the two loop corrections to the scalar mode
function \cite{BOW,KO}.}
\item{For a massless, minimally coupled scalar which is Yukawa-coupled
to a massless fermion --- the one loop corrections to the fermion
\cite{PW1,GP} and scalar \cite{DW} mode functions, and the two loop
vertex function \cite{MW1}.}
\item{For massless, minimally coupled scalar quantum electrodynamics
--- the one loop correction to the photon \cite{PTW,PW2} and scalar
\cite{KW1} mode functions, the two loop corrections to the coincident
scalar \cite{PTsW1} and field strength \cite{PTsW2} bilinears and
to the stress tensor \cite{PTsW2}.}
\item{For general relativity --- the one loop self-energy \cite{TW1}
and the expectation value of the graviton field at one \cite{LF,FMVV,TW2}
and two loops \cite{TW3}.}
\item{For general relativity coupled to a massless fermion ---
the one loop corrected fermion mode functions \cite{MW2}.}
\end{enumerate}
The leading secular effects of the first three models agree precisely
with the stochastic formalism of Starobinski\u{\i} \cite{AAS2,RPW3,TW4},
and the series of these leading effects at all orders can be resummed
\cite{SY,MW1,PTsW3} to give nonperturbative predictions. No stochastic
results are available for the last two models.

It is natural to extend these studies by combining general relativity
with a massless, minimally coupled scalar to probe the effects of
gravitons on scalars. Such an investigation has great phenomenological
interest because the inflaton potential of scalar-driven inflation is
so flat that inflaton mode functions are effectively those of a
massless, minimally coupled scalar. Indeed, it is standard to compute
the power spectrum of scalar perturbations by setting the scalar-scalar
correlator to its value for a massless, minimally coupled scalar on de
Sitter background \cite{MFB,LL}. The subject of quantum gravitational
corrections to the scalar effective potential has a long history
\cite{Lee,Linde} but we will here look at {\it all} one loop corrections
to the linearized, effective scalar field equation, including corrections
to the derivative terms and also the fully nonlocal corrections. This is
what one must do to fix the scalar field strength in addition to its mass.
It is worth noting that one loop quantum gravitational corrections
induce a time dependent field strength renormalization for massless
fermions which eventually becomes nonperturbatively large \cite{MW2}.

Our result is that the scalar mode functions experience no significant
corrections at one loop. We prove this by solving the linearized,
effective scalar field equation. Section 2 gives a brief review of our
previous one loop computation of the scalar self-mass-squared \cite{KW2},
which represents the quantum correction to the linearized effective
field equation. In section 3 we actually solve the equation in the
relevant regime of late times. Our conclusions comprise section 4.

\section{Our Previous Calculation}

The linearized effective scalar field equation consists of
the classical term minus the scalar self-mass-squared $M^2(x;x')$
integrated against the scalar,
\begin{equation}
\partial_{\mu} \Bigl(\sqrt{-g} g^{\mu\nu} \partial_{\nu} \varphi(x)\Bigr)
- \int d^4x' M^2(x;x') \varphi(x') = 0 \; .
\end{equation}
We computed $M^2(x;x')$ at one loop order using the bare Lagrangian of
Einstein + Scalar,
\begin{equation}
\mathcal{L} \equiv -\frac12 \partial_{\mu} \varphi \;
\partial_{\mu}\varphi g^{\mu\nu} \sqrt{-g} + \frac1{16\pi G}
\Bigl( R - (D\!-\!2) \Lambda \Bigr) \sqrt{-g} \; . \label{Lag}
\end{equation}
Here $\Lambda \equiv (D-1) H^2$ is the cosmological constant and
$G$ is Newton's constant. We worked on the open conformal coordinate
patch of $D$-dimensional de Sitter space, with the graviton field
$h_{\mu\nu}$ defined as,
\begin{equation}
g_{\mu\nu}(\eta,\vec{x}) \equiv a^2(\eta) \Bigl(\eta_{\mu\nu} +
\kappa h_{\mu\nu}(\eta,\vec{x}\Bigr) \qquad {\rm where} \qquad
a(\eta) \equiv -\frac1{H \eta} \; .
\end{equation}
Here $\eta_{\mu\nu}$ is the Minkowski metric and $\kappa^2 \equiv 16 \pi G$
is the loop counting parameter of quantum gravity. The graviton propagator
was computed by adding a non-de Sitter invariant gauge fixing term
\cite{TW5,RPW2},
\begin{eqnarray}
\mathcal{L}_{\rm GF} & \equiv & -\frac12 \sqrt{-g} g^{\mu\nu}
F_{\mu} F_{\nu} \; , \\
F_{\mu} & \equiv & \eta^{\rho\sigma} \Bigl( h_{\mu\rho , \sigma}
- \frac12 h_{\rho \sigma , \mu} + (D\!-\!2) a H h_{\mu \rho}
\delta^0_{\sigma} \Bigr) \; . \label{gauge}
\end{eqnarray}

The computation of $M^2(x;x')$ was made using dimensional regularization.
We subtracted off the one loop divergences using three dimension six
counterterms,
\begin{eqnarray}
\lefteqn{\Delta \mathcal{L} = \frac12 \alpha_1 \kappa^2 \varphi_{; \mu\nu}
\varphi_{; \rho\sigma} g^{\mu\nu} g^{\rho\sigma} \sqrt{-g} } \nonumber \\
& & \hspace{2cm} -\frac{\alpha_2 \kappa^2 R}{2 D (D\!-\!1)} \, \varphi_{, \mu}
\varphi_{, \nu} g^{\mu\nu} \sqrt{-g} - \frac{\alpha_3 \kappa^2 R}{
2 D (D\!-\!1)} \, \varphi_{, i} \varphi_{, j} g^{ij} \sqrt{-g} \; . \qquad
\end{eqnarray}
In these expressions a comma denotes ordinary differentiation while
a semi-colon stands for the covariant derivative. The noninvariant
counterterm proportional to $\alpha_3$ was made necessary by our
non-de Sitter invariant gauge fixing function (\ref{gauge}). Because
gravity is not perturbatively renormalizable, only the divergent parts
of the coefficients $\alpha_i$ are fixed,
\begin{eqnarray}
\alpha_1 & = & 0 + \Delta \alpha_1 \; , \\
\alpha_2 & = & \frac{H^{D-4}}{(4 \pi)^{\frac{D}2}} \Bigl\{ \frac{61}3\Bigr\}
+ \Delta \alpha_2 \; , \\
\alpha_3 & = & \frac{H^{D-4}}{(4\pi)^{\frac{D}2}} \Bigl\{ \frac4{D\!-\!4}
- \frac{58}3 - 2 \gamma\Bigr\} + \Delta \alpha_3 \; .
\end{eqnarray}
The finite parts $\Delta \alpha_i$ are arbitrary.

After renormalization and taking the unregulated limit ($D=4$), our
result is \cite{KW2},
\begin{equation}
-i M^2(x;x') = i \kappa^2 a^4 \Bigl(\Delta \alpha_1 \square^2 +
\Delta \alpha_2 \square + \Delta \alpha_3 \frac{\nabla^2}{a^2}\Bigr)
\delta^4(x \!-\! x') + \Bigl({\rm Table~\ref{FinNew}}\Bigr) \; . \label{final}
\end{equation}
The Laplacian $\nabla^2$ and the d'Alembertian $\square$ are,
\begin{equation}
\nabla^2 \equiv \partial_i \partial_i \qquad {\rm and} \qquad
\square \equiv \frac1{a^4} \partial^{\mu} \Bigl( a^2 \partial_{\mu}\Bigr)
= -\frac1{a^2} \partial_0^2 -\frac2{a} H \partial_0 + \frac1{a^2} \nabla^2 \; .
\end{equation}
We should also define the de Sitter invariant length function that appears
in Table~\ref{FinNew},
\begin{equation}
y(x;x') \equiv a(\eta) a(\eta') \Bigl\{ \Vert \vec{x} \!-\! \vec{x}' \Vert^2 -
(\vert \eta \!-\! \eta'\vert \!-\! i \delta)^2 \Bigr\} \; . \label{ydef}
\end{equation}

\begin{table}

\vbox{\tabskip=0pt \offinterlineskip
\def\tablerule{\noalign{\hrule}}
\halign to390pt {\strut#& \vrule#\tabskip=1em plus2em&
\hfil#\hfil& \vrule#& \hfil#\hfil& \vrule#\tabskip=0pt\cr
\tablerule \omit&height4pt&\omit&&\omit&\cr &&$\!\!\!\!{\rm
External \; Operator}\!\!\!\!$ && ${\rm Coefficient\ of}\; \frac{
\kappa^2 H^4}{(4\pi)^4}$ & \cr \omit&height4pt&\omit&&\omit&\cr
\tablerule \omit&height2pt&\omit&&\omit&\cr && $(aa')^4
\square^3/H^2$ && $-\frac{\ln{x}}{3x}$ & \cr
\omit&height2pt&\omit&&\omit&\cr \tablerule
\omit&height2pt&\omit&&\omit&\cr && $(aa')^4 \square^2$ &&
$\frac{26\ln{x}}{3x} +\frac{38}{3x} -6\ln^2{x} -18\ln{x}$ & \cr
\omit&height2pt&\omit&&\omit&\cr \tablerule
\omit&height2pt&\omit&&\omit&\cr && $(aa')^4 H^2\square$ &&
$-\frac{6\ln{x}}{x} + \frac{4}{x} -4 \ln{x}$ & \cr
\omit&height2pt&\omit&&\omit&\cr \tablerule
\omit&height2pt&\omit&&\omit&\cr && $(aa')^4 H^4$ &&
$\frac{4\ln{x}}{x}+\frac{18}{x}-108\ln{x} -120$ & \cr
\omit&height2pt&\omit&&\omit&\cr \tablerule
\omit&height2pt&\omit&&\omit&\cr && $(aa')^3(a^2+a'^2)
\square^3/H^2$ && $\frac{\ln{x}}{6x}$ & \cr
\omit&height2pt&\omit&&\omit&\cr \tablerule
\omit&height2pt&\omit&&\omit&\cr && $(aa')^3(a^2+a'^2) \square^2$
&& $-\frac{\ln{x}}{3x}+\frac{1}{6x}$ & \cr
\omit&height2pt&\omit&&\omit&\cr \tablerule
\omit&height2pt&\omit&&\omit&\cr && $(aa')^3(a^2+a'^2)H^2 \square$
&& $-\frac{2\ln{x}}{3x}+\frac{5}{x}-18\ln{x}$ & \cr
\omit&height2pt&\omit&&\omit&\cr \tablerule
\omit&height2pt&\omit&&\omit&\cr && $(aa')^3(a^2+a'^2)H^4$ &&
$\frac{4\ln{x}}{3x}-\frac{32}{3x} - 54$ & \cr
\omit&height2pt&\omit&&\omit&\cr \tablerule
\omit&height2pt&\omit&&\omit&\cr && $(aa')^3H^2\nabla^2$ &&
$-\frac{2\ln{x}}{3x}-\frac{16}{x} +84\ln{x}-48x\ln{x} + 96 x$ &
\cr \omit&height2pt&\omit&&\omit&\cr \tablerule
\omit&height2pt&\omit&&\omit&\cr && $(aa')^3 \nabla^2 \square$ &&
$\frac{\ln{x}}{3x}$ & \cr \omit&height2pt&\omit&&\omit&\cr
\tablerule \omit&height2pt&\omit&&\omit&\cr &&
$(aa')^2(a^2+a'^2)H^2\nabla^2$ && $-\frac{7\ln{x}}{3x} +
\frac{11}{2x} -12\ln{x}+48x\ln{x} + 12 x$ & \cr
\omit&height2pt&\omit&&\omit&\cr \tablerule
\omit&height2pt&\omit&&\omit&\cr &&
$(aa')^2(a^2+a'^2)\nabla^2\square$ &&
$\!\!\!\!\!-\frac{17\ln{x}}{6x} - \frac{49}{6x} + 4 \ln^2{x} + 10
\ln{x} + 12 x \ln{x}\!\!\!\!\!$ & \cr
\omit&height2pt&\omit&&\omit&\cr \tablerule
\omit&height2pt&\omit&&\omit&\cr && $(aa')^2\nabla^4$ &&
$\frac{10}{3}\ln{x} - 24 x \ln{x} + 24 x^2 \ln{x} - 36 x^2$ & \cr
\omit&height2pt&\omit&&\omit&\cr \tablerule}}

\caption{Nonlocal Contributions to $-iM^2(x;x')$. Here $x \equiv
\frac{y}{4}$ and $y(x;x')$ is defined in equation (\ref{ydef}).}

\label{FinNew}

\end{table}

\section{Effective Mode Equation}

This is the heart of the paper. We begin by clarifying what is
meant by the effective mode equation, then we explain the restricted
sense in which we solve it. Finally we work out the contributions
from the local counterterms in (\ref{final}) and from the nonlocal
terms of Table~\ref{FinNew}.

\subsection{The Schwinger-Keldysh Formalism} \label{sec:S-K}

We seek to find plane wave mode solutions to ``the effective field
equations.'' Although the quoted phrase is common parlance, it is
nonetheless ambiguous because there are different sorts of effective
field equations whose solutions mean different things in terms of
the unique canonical operator formalism. Introductory courses in
quantum field theory typically concern the in-out effective field
equations. The plane wave mode solutions of these equations give
in-out matrix elements of commutators of the full field with a
tree order creation operator of the in vacuum,
\begin{equation}
\Phi_{\rm io}(x;\vec{k}) = \Bigl\langle \Omega_{\rm out} \Bigl\vert
\Bigl[\varphi(x), \alpha^{\dagger}_{\rm in}(\vec{k}) \Bigr] \Bigr\vert
\Omega_{\rm in} \Bigr\rangle \; .
\end{equation}
This quantity is of great interest for flat space scattering
problems but it has little relevance to cosmology where there may
be an initial singularity and where particle production precludes
the in vacuum from evolving to the out vacuum.

The more interesting cosmological experiment is to release the
universe from a prepared state at finite time and let it evolve
as it will. The mode solutions of interest to this experiment
are the expectation values of commutators of the full field with the
tree order creation operator of the initial vacuum,
\begin{equation}
\Phi(x;\vec{k}) = \Bigl\langle \Omega \Bigl\vert \Bigl[\varphi(x),
\alpha^{\dagger}(\vec{k}) \Bigr] \Bigr\vert \Omega \Bigr\rangle \; .
\label{SKeqn}
\end{equation}
The effective field equation that $\Phi(x;\vec{k})$ obeys is given
by the Schwinger-Keldysh formalism \cite{MW2,KW1}. This is a covariant
covariant extension of Feynman diagrams which produces true expectation
values instead of in-out matrix elements \cite{JS,KTM,BM,LVK}. Because
there are excellent reviews on this subject \cite{RJ,CSHY,CH,FW}, we
will confine ourselves to explaining how to use the formalism.

The chief difference between the Schwinger-Keldysh and in-out
for\-mal\-isms is that the endpoints of particle lines have a $\pm$
polarity. Therefore, every propagator $i\Delta(x;x')$ of the in-out
formalism gives rise to four Schwinger-Keldysh propagators:
$i\Delta_{\scriptscriptstyle++}(x;x')$, $i\Delta_{\scriptscriptstyle
+-}(x;x')$, $i\Delta_{\scriptscriptstyle -+}(x;x')$ and $i\Delta_{
\scriptscriptstyle--}(x;x')$. Each of these propagators can be
obtained by making simple changes to the Feynman propagator. For our
model, the Feynman propagators of the scalar and graviton happen to
depend upon the length function $y(x;x')$ defined in expression
(\ref{ydef}), and also upon the two scale factors. The four polarities
derive from making the following substitutions for $y(x;x')$:
\begin{eqnarray}
i\Delta_{\scriptscriptstyle ++}(x;x')\, : & & y \rightarrow
y_{\scriptscriptstyle ++}(x;x') \equiv a(\eta) a(\eta') \Bigl[ \Vert \vec{x}
\!-\! \vec{x}' \Vert^2 - (\vert \eta\!-\! \eta'\vert \!-\! i \delta )^2 \Bigr]
\; , \label{y++} \qquad \\
i\Delta_{\scriptscriptstyle +-}(x;x')\, : & & y \rightarrow
y_{\scriptscriptstyle +-}(x;x') \equiv a(\eta) a(\eta') \Bigl[ \Vert \vec{x}
\!-\! \vec{x}' \Vert^2 - (\eta\!-\! \eta' \!+\! i \delta )^2 \Bigr]
\; , \label{y+-} \qquad \\
i\Delta_{\scriptscriptstyle -+}(x;x')\, : & & y \rightarrow
y_{\scriptscriptstyle -+}(x;x') \equiv a(\eta) a(\eta') \Bigl[ \Vert \vec{x}
\!-\! \vec{x}' \Vert^2 - (\eta\!-\! \eta' \!-\! i \delta )^2 \Bigr] \; ,
\qquad \\
i\Delta_{\scriptscriptstyle --}(x;x')\, : & & y \rightarrow
y_{\scriptscriptstyle --}(x;x') \equiv a(\eta) a(\eta') \Bigl[ \Vert \vec{x}
\!-\! \vec{x}' \Vert^2 - (\vert \eta\!-\! \eta'\vert \!+\! i \delta )^2 \Bigr]
\; . \qquad
\end{eqnarray}
Vertices in the Schwinger-Keldysh formalism either have all $+$ lines
or all $-$ lines. The $+$ vertex is identical to that of the in-out
formalism, whereas the $-$ vertex is its conjugate.

Because any external line can be either $+$ or $-$, each N-point
one particle irreducible (1PI) function of the in-out formalism gives
rise to $2^N$ 1PI functions in the Schwinger-Keldysh formalism. The
Schwinger-Keldysh effective action is the generating functional of
these 1PI functions. We can express it in terms of fields $\varphi_+$,
to access the $+$ lines, and $\varphi_-$, to access the $-$ lines,
\begin{eqnarray}
\lefteqn{\Gamma[\varphi_+,\varphi_-] = S[\varphi_+] - S[\varphi_-]
- \frac12 \int \! d^4x \! \int \! d^4x' } \nonumber \\
& & \left\{ \matrix{ \varphi_+(x)
M^2_{ \scriptscriptstyle ++}(x;x') \varphi_+(x') + \varphi_+(x)
M^2_{\scriptscriptstyle +-}(x;x') \varphi_-(x') \cr +\varphi_-(x)
M^2_{\scriptscriptstyle -+}(x;x') \varphi_+(x') + \varphi_-(x)
M^2_{\scriptscriptstyle --}(x;x') \varphi_-(x') } \right\} \!+\!
O(\varphi^3_{\pm}) , \qquad
\end{eqnarray}
where $S[\varphi]$ is the classical scalar action.

At the order we are working, $-iM^2_{\scriptscriptstyle ++}(x;x')$ is
the same as the in-out self-mass-squared. We can therefore read it off
from (\ref{final}). We get $-i M^2_{\scriptscriptstyle +-}(x;x')$ by
dropping the delta function terms, reversing the sign and replacing
$y(x;x')$ by $y_{\scriptscriptstyle +-}(x;x')$ in Table~\ref{FinNew}.
The other two 1PI 2-point functions derive from conjugating these two,
\begin{equation}
-i M^2_{\scriptscriptstyle --}(x;x') = \Bigl(-i
M^{2}_{\scriptscriptstyle ++}( x;x') \Bigr)^* \; , -i
M^2_{\scriptscriptstyle -+}(x;x') = \Bigl( -i
M^2_{\scriptscriptstyle +-}(x;x')\Bigr)^* .
\end{equation}

To get the Schwinger-Keldysh effective field equations one varies the
action with respect to the field of either polarity, then sets the two
polarities equal to $\Phi(x)$. At linearized level this gives,
\begin{equation}
a^4 \square \Phi(x) - \int_{\eta_i}^0 \! d\eta'\!\int\!d^3x' \,
\Bigl\{M^2_{\scriptscriptstyle ++}(x;x') + M^2_{\scriptscriptstyle +-}(x;x')
\Bigr\} \Phi(x') = 0 \; . \qquad \label{lineqn}
\end{equation}
Here $\eta_i = -1/H$ is the initial (conformal) time at which the
universe is released in free Bunch-Davies vacuum. One can see from
relations (\ref{y++}) and (\ref{y+-}), and from the extra conjugated
vertex in $M_{\scriptscriptstyle +-}(x;x')$, that the bracketed term
vanishes for $\eta' > \eta$. The fact that $y_{\scriptscriptstyle +-}(x;x')$
is the complex conjugate of $y_{\scriptscriptstyle ++}(x;x')$ for $\eta'
< \eta$ means that the bracketed term is real. One also sees that it
must involve the imaginary part of at least one propagator, which means
the only net effect comes from points $x^{\prime \mu}$ on or inside the
past light-cone of $x^{\mu}$. Hence the Schwinger-Keldysh effective field
equations are real and causal, unlike those of the in-out formalism.

\subsection{Restrictions on Our Solution} \label{sec:solve}

Two limitations on our knowledge impose important restrictions on the
sense in which we can solve (\ref{lineqn}):
\begin{enumerate}
\item{We only know the scalar self-mass-squared at one loop order; and}
\item{We took the initial state to be free, Bunch-Davies vacuum.}
\end{enumerate}
The first limitation means we must solve (\ref{lineqn}) perturbatively.
The full scalar self-mass-squared can be expanded in powers of
the loop-counting parameter $\kappa^2 = 16\pi G$,
\begin{equation}
M^2_{\scriptscriptstyle ++}(x;x') + M^2_{\scriptscriptstyle +-}(x;x')
= \sum_{\ell=1}^{\infty} \kappa^{2\ell} \mathcal{M}^2_{\ell}(x;x') \; .
\label{mass}
\end{equation}
A similar expansion applies for plane wave solutions to (\ref{lineqn}),
\begin{equation}
\Phi(x;\vec{k}) = \sum_{\ell=0}^{\infty} \kappa^{2 \ell} \Phi_{\ell}(\eta,k)
\times e^{i \vec{k} \cdot \vec{x}} \; .
\end{equation}
To make $\Phi(x;\vec{k})$ agree with (\ref{SKeqn}) we must normalize
the tree order solution appropriately,
\begin{equation}
\Phi_0(\eta,k) = u(\eta,k) \equiv \frac{H}{\sqrt{2 k^3}} \Bigl(1 -
\frac{i k}{a H}\Bigr) \exp\Bigl[ \frac{ik}{a H}\Bigr] \; .
\end{equation}
The $\ell \ge 1$ solutions obey,
\begin{equation}
a^2 \Bigl[\partial_0^2 + 2 H \partial_0 + k^2\Bigr] \Phi_{\ell}(\eta,k)
= -\!\sum_{k=1}^{\ell} \!\int_{\eta_i}^0 \!\!\! d\eta' \!\! \int \!\!
d^3x' \mathcal{M}^2_k(x;x') \Phi_{\ell-k}(\eta',k) e^{i \vec{k} \cdot (\vec{x}'
-\vec{x})} \!\! . \label{elleqn}
\end{equation}
We know only $\mathcal{M}^2_1(x;x')$ so the sole correction we can compute
is $\Phi_1(\eta,k)$.

The second limitation means it only makes sense to solve for
$\Phi_1(\eta,k)$ at late times, i.e., as $\eta \rightarrow 0^-$.
Interactions result in important corrections to free vacuum on a flat
background and it is unthinkable that this does not happen as well
for de Sitter background. In the Schwinger-Keldysh formalism these
corrections would correspond to vertices on the initial value
surface \cite{LF,KW1}. In the in-out formalism the free vacuum is
automatically corrected by time evolution. One can follow the progress
of this in the Schwinger-Keldysh formalism by isolating terms that
decay with increasing time after the release of the initial state.
For example, the two loop expectation value of the stress tensor of
a massless, minimally coupled scalar with a quartic self-interaction
gives the following energy density and pressure \cite{OW},
\begin{eqnarray}
\rho & = & \frac{\lambda H^4}{(2 \pi)^4} \Biggl\{\frac18 \ln^2(a) \!+\!
\frac1{18 a^3} \!-\! \frac18 \sum_{n=1}^{\infty} \frac{(n\!+\!2) a^{-n-1}}{
(n\!+\!1)^2} \Biggr\} + O(\lambda^2) \; , \\
p & = & \frac{\lambda H^4}{(2 \pi)^4} \Biggl\{-\frac18 \ln^2(a) \!-\!
\frac1{12} \ln(a) \!-\! \frac1{24} \sum_{n=1}^{\infty} \frac{(n^2\!-\!4)
a^{-n-1}}{(n\!+\!1)^2} \Biggr\} + O(\lambda^2) \; . \qquad
\end{eqnarray}
We suspect that the (separately conserved) terms which fall like powers
of $1/a$ can be absorbed into an order $\lambda$ correction of the initial
state. On the other hand, the terms which grow like powers of $\ln(a)$
represent the effect of inflationary particle production (in this case,
of scalars) pushing the field up its quartic potential.

Because we have not worked out the order $\kappa$ and $\kappa^2$
corrections to Bunch-Davies vacuum, we cannot trust corrections to
$\Phi_1(\eta,k)$ that fall off at late times relative to the tree
order solution $u(\eta,k)$,
\begin{equation}
u(\eta,k) = \frac{H}{\sqrt{2 k^3}} \Bigl\{ 1 + \frac{k^2}{2 H^2 a^2}
+ \frac{i k^3}{3 H^3 a^3} + O\Bigl(\frac{k^4}{H^4 a^4}\Bigr) \Bigr\} \; .
\label{asymp}
\end{equation}
To understand what this means, it is best to convert equation (\ref{elleqn})
for $\Phi_1$ from conformal time $\eta$ to comoving time $t \equiv
-\ln(-H \eta)/H$,
\begin{equation}
\Bigl[ \frac{\partial^2}{\partial t^2} + 3 H \frac{\partial}{\partial t}
+ \frac{k^2}{a^2}\Bigr] \Phi_1 = -\frac1{a^4} \!\int_{\eta_i}^0 \!\!\!
d\eta' \, u(\eta',k) \!\! \int \!\! d^3x' \mathcal{M}^2_1(x;x') e^{i \vec{k}
\cdot (\vec{x}' -\vec{x})} . \label{keyeqn}
\end{equation}
At late times we can drop the factor of $k^2/a^2$ on the left hand side.
Given a putative form for the late time behavior of the right hand side
it is easy to infer the leading late time behavior of $\Phi_1$, for
example,
\begin{eqnarray}
{\rm r.h.s} \longrightarrow \ln(a) & \Longrightarrow & \Phi_1
\longrightarrow \frac{\ln^2(a)}{6 H^2} \; , \\
{\rm r.h.s} \longrightarrow 1 & \Longrightarrow & \Phi_1 \longrightarrow
\frac{\ln(a)}{3 H^2} \; , \\
{\rm r.h.s} \longrightarrow \frac{\ln(a)}{a} & \Longrightarrow & \Phi_1
\longrightarrow {\rm Constant} -\frac{\ln(a)}{2 H^2 a} \; , \\
{\rm r.h.s} \longrightarrow \frac1{a} & \Longrightarrow & \Phi_1
\longrightarrow {\rm Constant} -\frac1{2 H^2 a} \; .
\end{eqnarray}
Because any constant contribution to $\Phi_1$ could be absorbed into a
field strength renormalization, the only effects which can be distinguished
from corrections to the initial state derive from contributions to the
right hand side that fall off no faster than $1/\ln(a)$. It turns out
that inverse powers of $\ln(a)$ cannot occur, so the practical dividing
line is between contributions to the right hand side which grow or
approach a nonzero constant and those which fall off.

\subsection{Local Corrections}

Because the scalar d'Alembertian annihilates the tree order solution,
only the third, noncovariant counterterm makes any contribution to
(\ref{keyeqn}),
\begin{eqnarray}
\lefteqn{\frac1{a^4} \int d^4x' \, a^4 \Bigl(\Delta \alpha_1 \square^2
+ \Delta \alpha_2 \square + \Delta \alpha_3 \frac{\nabla^2}{a^2} \Bigr)
\delta^4(x \!-\! x') } \nonumber \\
& & \hspace{5.3cm} \times u(\eta',\vec{k}) e^{i \vec{k} \cdot (\vec{x}' -
\vec{x})} = -\Delta \alpha_3 \frac{k^2}{a^2} \, u(\eta,k) \; . \qquad
\end{eqnarray}
This term rapidly redshifts to zero and we see from the preceding
discussion that it cannot make a significant contribution to $\Phi_1$
at late times.

\subsection{Nonlocal Corrections}

Although Table~\ref{FinNew} might seem to present a bewildering
variety of nonlocal contributions to (\ref{keyeqn}), a series of
seven straightforward steps suffices to evaluate each one:
\begin{enumerate}
\item{Eliminate any factors of $1/y$ using the identities,
\begin{eqnarray}
\frac4{y} & = & \frac{\square}{H^2} \Biggl\{
\ln\Bigl(\frac{y}4\Bigr)\Biggr\} + 3 \; , \label{1/x} \qquad \\
\frac4{y} \ln\Bigl(\frac{y}4\Bigr) & = & \frac{\square}{H^2}
\Biggl\{ \frac12 \ln^2\Bigl(\frac{y}4\Bigr) \!-\!
\ln\Bigl(\frac{y}4\Bigr) \Biggr\} + 3\ln\Bigl(\frac{y}4\Bigr) - 2
\; . \label{ln/x} \qquad
\end{eqnarray}}
\item{Extract the factors of $\square$ and $\nabla^2$ from the
integration over $x^{\prime \mu}$ using the identities,
\begin{eqnarray}
\frac{\square}{H^2} & \longrightarrow & -\Bigl[ a^2 \frac{\partial^2}{
\partial a^2} + 4 a \frac{\partial}{\partial a} + \frac{k^2}{a^2 H^2}
\Bigr] \; , \label{box} \\
\nabla^2 & \longrightarrow & - k^2 \; .
\end{eqnarray}}
\item{Combine the $++$ and $+-$ terms to extract a factor of $i$
and make causality manifest,
\begin{eqnarray}
\ln\Bigl(\frac{y_{\scriptscriptstyle ++}}4\Bigr) -
\ln\Bigl(\frac{y_{\scriptscriptstyle +-}}4\Bigr) & \!\!\!\!=\!\!\!\! &
2 \pi i \theta\Bigl(\Delta \eta \!-\! \Delta x \Bigr) \; , \qquad \\
\ln^2\!\Bigl(\frac{y_{\scriptscriptstyle ++}}4\!\Bigr) - \ln^2\!\Bigl(\frac{
y_{\scriptscriptstyle +-}}4\!\Bigr) & \!\!\!\!=\!\!\!\! & 4 \pi i
\theta\Bigl(\Delta \eta \!-\! \Delta x\Bigr) \ln\Bigl(\frac14 a a'
H^2 (\Delta \eta^2 \!-\! \Delta x^2) \Bigr) \; . \qquad
\end{eqnarray}
Here we define $a \equiv a(\eta)$, $a' \equiv a(\eta')$,
$\Delta \eta \equiv \eta - \eta'$ and $\Delta x \equiv \Vert \vec{x}
- \vec{x}'\Vert$. Note that any positive powers of $y$ become,
\begin{equation}
y_{\scriptscriptstyle +\pm} \longrightarrow -\frac14 a a' H^2
(\Delta \eta^2 \!-\! \Delta x^2) \; .
\end{equation}}
\item{Make the change of variables $\vec{r} = \vec{x}' - \vec{x}$,
perform the angular integrations and make the further change of
variable $r = \Delta \eta \cdot z$,
\begin{eqnarray}
\lefteqn{\int \! d^3x' \theta\Bigl(\Delta \eta \!-\! \Delta x\Bigr)
F\Bigl(\frac14 a a' H^2 (\Delta \eta^2 \!-\! \Delta x^2)\Bigr)
e^{i \vec{k} \cdot (\vec{x}' - \vec{x})} } \nonumber \\
& & = 4\pi \theta(\Delta \eta) \int_0^{\Delta \eta}
\!\! dr \, r^2 F\Bigl(\frac14 a a' H^2 (\Delta \eta^2 \!-\! r^2)\Bigr)
\frac{\sin(k \Delta x)}{k \Delta x} \; , \\
& & = 4\pi \theta(\Delta \eta) \Delta \eta^3 \int_0^1 \!\! dz \,
z^2 F\Biggl(a a' \Bigl(\frac1{a'} \!-\! \frac1{a}\Bigr)^2 \Bigl(\frac{1
\!-\! z^2}4\Bigr) \Biggr) \frac{\sin(k \Delta \eta z)}{k \Delta \eta z}
\; . \label{ang} \qquad
\end{eqnarray}}
\item{Reduce the $z$ integration to a combination of elementary
functions and sine and cosine integrals \cite{DW}.}
\item{Make the change of variables $a' = -1/H\eta'$, expand the integrand
and perform the integration over $a'$.}
\item{Act any derivatives with respect to $a$.}
\end{enumerate}

Much of the labor involved in implementing these steps derives from the
spacetime dependence of the zeroth order solution, $u(\eta',k) e^{i\vec{k}
\cdot \vec{x}'}$. For example, one can see from (\ref{ang}) that only
elementary functions would result from the $z$ integration if the zeroth
order solution were constant. In fact it is completely justified to
make this simplification,
\begin{eqnarray}
\lefteqn{-\frac1{a^4} \int_{\eta_i}^{0} \!\!\! d\eta' \!\!\! \int \!\! d^3x'
\mathcal{M}^2_1(x;x') u(\eta',k) e^{i \vec{k} \cdot (\vec{x}' -
\vec{x})} } \nonumber \\
& & \hspace{4.5cm} \longrightarrow - \frac{u(0,k)}{a^4} \!\! \int_{\eta_i}^{0}
\!\!\! d\eta' \!\!\! \int \!\! d^3x' \mathcal{M}^2_1(x;x') \, ; \qquad
\label{replace}
\end{eqnarray}
To see why, note from expressions (\ref{asymp}) and (\ref{ang}) that the
deviation of the zeroth order solution from $u(0,k)$ introduces at least
two factors of $1/a'$ or $1/a$. Because the constant mode function
can at best result in powers of $\ln(a)$, and because it makes no
physical sense to retain contributions to (\ref{keyeqn}) which fall off,
we can immediately discard terms which acquire an extra factor of $1/a$.
Extra factors of $1/a'$ effectively restrict the integration to early
times, which again causes the net result to fall off at late times.

One consequence of the simplification (\ref{replace}) is that we can
neglect any contribution from Table~\ref{FinNew} which contains a
factor of $\nabla^2$. We therefore need only compute terms of the form,
\begin{eqnarray}
-u(0,k) \frac{i H^8 a^{4-K}}{(4 \pi)^4}
\Bigl(\frac{\square}{H^2}\Bigr)^N \!\! \int_{\eta_i}^{0} \!\!
d\eta' \, a^{\prime K} \!\! \int \!\! d^3x' \Biggl\{
f\Bigl(\frac{y_{\scriptscriptstyle ++}}4\Bigr) \!-\!
f\Bigl(\frac{y_{\scriptscriptstyle +-}}4\Bigr) \Biggr\} \; , \label{eight}
\end{eqnarray}
where the constant $K$ takes the values of 3, 4 and 5, and the functions
$f(x)$ are $1/x$, $\ln{x}/x$, $\ln{x}$ and $\ln^2{x}$. The action of the
d'Alembertian derives from setting $k=0$ in (\ref{box}). The integrations
for $K = 3$ (given in Table~\ref{f3int}) and $K = 4$ (given in
Table~\ref{f4int}) were worked out in a previous paper \cite{PTsW1}.
Applying the same technique --- which is just the 7-step procedure given
above for $k=0$ --- gives the results for $K=5$ in Table~\ref{f5int}.

\begin{table}
\vbox{\tabskip=0pt \offinterlineskip
\def\tablerule{\noalign{\hrule}}
\halign to390pt {\strut#& \vrule#\tabskip=1em plus2em&
\hfil#\hfil& \vrule#& \hfil#\hfil& \vrule#\tabskip=0pt\cr
\tablerule \omit&height4pt&\omit&&\omit&\cr
\omit&height2pt&\omit&&\omit&\cr && $\!\!\!\!\! f(x) \!\!\!\!\!$
&& $\!\!\!\!\! -\frac{i H^4}{16 \pi^2} \times \int d^4x' a^{\prime 3}
\{f(\frac{y_{++}}{4}) - f(\frac{y_{+-}}{4})\} \!\!\!\!\!$ & \cr
\omit&height4pt&\omit&&\omit&\cr \tablerule
\omit&height2pt&\omit&&\omit&\cr \tablerule
\omit&height2pt&\omit&&\omit&\cr && $\!\!\!\!\! \frac1{x}
\!\!\!\!\!$ && $\!\!\!\!\! -\frac{\ln(a)}{a} + \frac1{a}
+ O(\frac1{a^2}) \!\!\!\!\!$ & \cr
\omit&height2pt&\omit&&\omit&\cr \tablerule
\omit&height2pt&\omit&&\omit&\cr && $\!\!\!\!\! \frac{\ln(x)}{x}
\!\!\!\!\!$ && $\!\!\!\!\! -\frac{\ln^2(a)}{2 a} + \frac{2 \ln(a)}{a}
- \frac{3}{a} + \frac{\pi^2}{3 a} + O(\frac{\ln(a)}{a^2}) \!\!\!\!\!$ & \cr
\omit&height2pt&\omit&&\omit&\cr \tablerule
\omit&height2pt&\omit&&\omit&\cr && $\!\!\!\!\! {\scriptstyle \ln(x)}
\!\!\!\!\!$ && $\!\!\!\!\! \frac16 - \frac{\ln(a)}{2 a} + \frac1{4 a}
+ O(\frac1{a^2}) \!\!\!\!\!$ & \cr
\omit&height2pt&\omit&&\omit&\cr \tablerule
\omit&height2pt&\omit&&\omit&\cr && $\!\!\!\!\! {\scriptstyle \ln^2(x)}
\!\!\!\!\!$ && $\!\!\!\!\! \frac13 {\scriptstyle \ln(a)} \!-\! \frac{11}9
\!-\! \frac{\ln^2(a)}{2 a} \!+\! \frac{2 \ln(a)}{a} \!-\! \frac{9}{4 a}
\!+\! \frac{\pi^2}{3 a} \!+\! O(\frac{\ln(a)}{a^2}) \!\!\!\!\!$ & \cr
\omit&height2pt&\omit&&\omit&\cr \tablerule
\omit&height2pt&\omit&&\omit&\cr \tablerule}}

\caption{Integrals with $a^{\prime 3}$.}

\label{f3int}

\end{table}

\begin{table}
\vbox{\tabskip=0pt \offinterlineskip
\def\tablerule{\noalign{\hrule}}
\halign to370pt {\strut#& \vrule#\tabskip=1em plus2em&
\hfil#\hfil& \vrule#& \hfil#\hfil& \vrule#\tabskip=0pt\cr
\tablerule \omit&height4pt&\omit&&\omit&\cr
\omit&height2pt&\omit&&\omit&\cr && $\!\!\!\!\! f(x) \!\!\!\!\!$
&& $\!\!\!\!\! -\frac{i H^4}{16 \pi^2} \times \int d^4x' a^{\prime 4}
\{f(\frac{y_{++}}{4}) - f(\frac{y_{+-}}{4})\} \!\!\!\!\!$ & \cr
\omit&height4pt&\omit&&\omit&\cr \tablerule
\omit&height2pt&\omit&&\omit&\cr \tablerule
\omit&height2pt&\omit&&\omit&\cr && $\!\!\!\!\! \frac1{x}
\!\!\!\!\!$ && $\!\!\!\!\! - \frac12 + O(\frac1{a}) \!\!\!\!\!$ &
\cr \omit&height2pt&\omit&&\omit&\cr \tablerule
\omit&height2pt&\omit&&\omit&\cr && $\!\!\!\!\! \frac{\ln(x)}{x}
\!\!\!\!\!$ && $\!\!\!\!\! \frac34 + O(\frac1{a}) \!\!\!\!\!$ &
\cr \omit&height2pt&\omit&&\omit&\cr \tablerule
\omit&height2pt&\omit&&\omit&\cr && $\!\!\!\!\! {\scriptstyle
\ln(x)} \!\!\!\!\!$ && $\!\!\!\!\! \frac16 {\scriptstyle \ln(a)} -
\frac{11}{36} + O(\frac1{a}) \!\!\!\!\!$ & \cr
\omit&height2pt&\omit&&\omit&\cr \tablerule
\omit&height2pt&\omit&&\omit&\cr && $\!\!\!\!\! {\scriptstyle
\ln^2(x)} \!\!\!\!\!$ && $\!\!\!\!\! \frac16 {\scriptstyle
\ln^2(a)} - \frac89 {\scriptstyle \ln(a)} + \frac74 -
\frac{\pi^2}{9} + O(\frac{\ln(a)}{a}) \!\!\!\!\!$ & \cr
\omit&height2pt&\omit&&\omit&\cr \tablerule
\omit&height2pt&\omit&&\omit&\cr \tablerule}}

\caption{Integrals with $a^{\prime 4}$.}

\label{f4int}

\end{table}

\begin{table}
\vbox{\tabskip=0pt \offinterlineskip
\def\tablerule{\noalign{\hrule}}
\halign to370pt {\strut#& \vrule#\tabskip=1em plus2em&
\hfil#\hfil& \vrule#& \hfil#\hfil& \vrule#\tabskip=0pt\cr
\tablerule \omit&height4pt&\omit&&\omit&\cr
\omit&height2pt&\omit&&\omit&\cr && $\!\!\!\!\! f(x) \!\!\!\!\!$
&& $\!\!\!\!\! -\frac{i H^4}{16 \pi^2} \times \int d^4x' a^{\prime 5}
\{f(\frac{y_{++}}{4}) - f(\frac{y_{+-}}{4})\} \!\!\!\!\!$ & \cr
\omit&height4pt&\omit&&\omit&\cr \tablerule
\omit&height2pt&\omit&&\omit&\cr \tablerule
\omit&height2pt&\omit&&\omit&\cr
&& $\!\!\!\!\! \frac1{x} \!\!\!\!\!$ &&
$\!\!\!\!\! -{\scriptstyle \frac16 a} + O(\frac1{a}) \!\!\!\!\!$ & \cr
\omit&height2pt&\omit&&\omit&\cr
\tablerule \omit&height2pt&\omit&&\omit&\cr
&& $\!\!\!\!\! \frac{\ln(x)}{x} \!\!\!\!\!$ && $\!\!\!\!\! {\scriptstyle
\frac{17}{36} a} + O(\frac1{a}) \!\!\!\!\!$ & \cr
\omit&height2pt&\omit&&\omit&\cr \tablerule
\omit&height2pt&\omit&&\omit&\cr
&& $\!\!\!\!\! {\scriptstyle \ln(x)} \!\!\!\!\!$ && $\!\!\!\!\!
\frac1{24} {\scriptstyle a} -\frac16 + O(\frac1{a}) \!\!\!\!\!$ & \cr
\omit&height2pt&\omit&&\omit&\cr \tablerule
\omit&height2pt&\omit&&\omit&\cr
&& $\!\!\!\!\! {\scriptstyle \ln^2(x)} \!\!\!\!\!$
&& $\!\!\!\!\! -\frac{13}{144} {\scriptstyle a} \!-\! \frac13
{\scriptstyle \ln(a)} \!+\! \frac5{9} \!+\! O(\frac{\ln(a)}{a}) \!\!\!\!\!$
& \cr
\omit&height2pt&\omit&&\omit&\cr \tablerule
\omit&height2pt&\omit&&\omit&\cr \tablerule}}

\caption{Integrals with $a^{\prime 5}$.}

\label{f5int}

\end{table}

It remains just to act the d'Alembertians and sum the results for
each of the contributions from Table~\ref{FinNew}. That is done in
Table~\ref{FinTot}. Although individual contributions can grow as
fast as $\ln^2(a)$, all growing or even finite terms cancel. It
follows that the right hand side of (\ref{keyeqn}) falls off at
least as fast as $1/a$ times powers of $\ln(a)$. Hence there are
no significant corrections to the mode function at one loop order.

\begin{table}

\vbox{\tabskip=0pt \offinterlineskip
\def\tablerule{\noalign{\hrule}}
\halign to390pt {\strut#& \vrule#\tabskip=1em plus2em&
\hfil#\hfil& \vrule#& \hfil#\hfil& \vrule#\tabskip=0pt\cr
\tablerule \omit&height4pt&\omit&&\omit&\cr &&$\!\!\!\!{\rm
External \; Operator} \times f(x)\!\!\!\!$ && ${\rm Coefficient\
of}\; u(0,k) \times \frac{H^4}{16 \pi^2}$ & \cr
\omit&height4pt&\omit&&\omit&\cr \tablerule
\omit&height2pt&\omit&&\omit&\cr && $(aa')^4 \square^3/H^2 \times
-\frac{\ln{x}}{3x}$ && $0$ & \cr \omit&height2pt&\omit&&\omit&\cr
\tablerule \omit&height2pt&\omit&&\omit&\cr && $(aa')^4 \square^2
\times \frac{26\ln{x}}{3x}$ && $ 0$ & \cr
\omit&height2pt&\omit&&\omit&\cr \tablerule
\omit&height2pt&\omit&&\omit&\cr && $(aa')^4 \square^2 \times
\frac{38}{3x}$ && $0$ & \cr \omit&height2pt&\omit&&\omit&\cr
\tablerule \omit&height2pt&\omit&&\omit&\cr && $(aa')^4 \square^2
\times -6\ln^2x$ && $-18 $ & \cr \omit&height2pt&\omit&&\omit&\cr
\tablerule \omit&height2pt&\omit&&\omit&\cr && $(aa')^4 \square^2
\times -18\ln{x}$ && $0$ & \cr \omit&height2pt&\omit&&\omit&\cr
\tablerule \omit&height2pt&\omit&&\omit&\cr && $(aa')^4 H^2\square
\times -\frac{6\ln{x}}{x}$ && $ 0$ & \cr
\omit&height2pt&\omit&&\omit&\cr \tablerule
\omit&height2pt&\omit&&\omit&\cr && $(aa')^4 H^2\square \times
\frac{4}{x}$ && $0$ & \cr \omit&height2pt&\omit&&\omit&\cr
\tablerule \omit&height2pt&\omit&&\omit&\cr && $(aa')^4 H^2\square
\times -4 \ln{x}$ && $2$ & \cr \omit&height2pt&\omit&&\omit&\cr
\tablerule \omit&height2pt&\omit&&\omit&\cr && $(aa')^4 H^4 \times
\frac{4\ln{x}}{x}$ && $3 $ & \cr \omit&height2pt&\omit&&\omit&\cr
\tablerule \omit&height2pt&\omit&&\omit&\cr && $(aa')^4 H^4 \times
\frac{18}{x}$ && $-9 $ & \cr \omit&height2pt&\omit&&\omit&\cr
\tablerule \omit&height2pt&\omit&&\omit&\cr && $(aa')^4 H^4 \times
-108\ln{x}$ && $33 \; -18 \;  \ln(a)$ & \cr
\omit&height2pt&\omit&&\omit&\cr \tablerule
\omit&height2pt&\omit&&\omit&\cr && $(aa')^3(a^2+a'^2)
\square^3/H^2 \times \frac{\ln{x}}{6x}$ && $(\frac{-325 +
12\pi^2}{27}) + \frac{14}{3} \ln(a) -\frac23 \ln^2(a)$ & \cr
\omit&height2pt&\omit&&\omit&\cr \tablerule
\omit&height2pt&\omit&&\omit&\cr && $(aa')^3(a^2+a'^2) \square^2
\times -\frac{\ln{x}}{3x}$ && $(\frac{85 - 12\pi^2}{27}) -4 \ln(a)
+\frac23 \ln^2(a)$ & \cr \omit&height2pt&\omit&&\omit&\cr
\tablerule \omit&height2pt&\omit&&\omit&\cr && $(aa')^3(a^2+a'^2)
\square^2 \times \frac{1}{6x}$ && $\frac89 \; -\frac23 \ln(a)$ &
\cr \omit&height2pt&\omit&&\omit&\cr \tablerule
\omit&height2pt&\omit&&\omit&\cr && $(aa')^3(a^2+a'^2)H^2 \square
\times -\frac{2\ln{x}}{3x}$ && $(\frac{160 - 12\pi^2}{27})
-\frac{10}{3} \ln(a) +\frac23 \ln^2(a)$ & \cr
\omit&height2pt&\omit&&\omit&\cr \tablerule
\omit&height2pt&\omit&&\omit&\cr && $(aa')^3(a^2+a'^2)H^2 \square
\times \frac{5}{x}$ && $\frac{55}{3} \; -10 \ln(a) $ & \cr
\omit&height2pt&\omit&&\omit&\cr \tablerule
\omit&height2pt&\omit&&\omit&\cr && $(aa')^3(a^2+a'^2)H^2 \square
\times -18\ln{x}$ && $-15 \;+18 \ln(a)$ & \cr
\omit&height2pt&\omit&&\omit&\cr \tablerule
\omit&height2pt&\omit&&\omit&\cr && $(aa')^3(a^2+a'^2)H^4 \times
\frac{4\ln{x}}{3x}$ && $(\frac{-91 + 12 \pi^2}{27}) +\frac{8}{3}
\ln(a) -\frac23 \ln^2(a)$ & \cr \omit&height2pt&\omit&&\omit&\cr
\tablerule \omit&height2pt&\omit&&\omit&\cr &&
$(aa')^3(a^2+a'^2)H^4 \times -\frac{32}{3x}$ && $-\frac{80}{9} +
\frac{32}{3} \ln(a)$ & \cr \omit&height2pt&\omit&&\omit&\cr
\tablerule \omit&height2pt&\omit&&\omit&\cr \tablerule
\omit&height2pt&\omit&&\omit&\cr && Total && $0$ & \cr
\omit&height2pt&\omit&&\omit&\cr \tablerule}}

\caption{$-\frac{u(0,k)}{a^4} \frac{i H^8}{(4\pi)^4} \int d^4x' \,
({\rm Ext.\ Operator}) \times \{f(\frac{y_{++}}4) - f(\frac{y_{+-}}4)\}$
for each contributing nonlocal term.}

\label{FinTot}

\end{table}

\section{Discussion}

We have solved the one loop-corrected, linearized effective field
equation for a massless, minimally coupled scalar interacting with
gravity during de Sitter inflation. Unlike previous analyses of
the scalar effective potential \cite{Linde}, our technique uses
the full self-mass-squared, including corrections to the
derivative terms and nonlocal corrections. It should therefore be
sensitive not only to the scalar's mass but also to its field
strength. We find no significant corrections of any sort at one
loop order.

Physically our result means that the sea of infrared gravitons
produced by inflation has little effect on the scalar. Although our
scalar is a spectator to $\Lambda$-driven inflation, typical scalar
inflaton potentials are so flat that the inflaton is also unlikely
to suffer significant one loop corrections from quantum gravity. If
so, it is consistent to use the tree order scalar mode functions to
compute the power spectrum of cosmological perturbations.

There are good reasons why the scalar should not acquire a mass
from quantum gravity \cite{Linde} but there seems to be no reason
why its field strength cannot suffer a time dependent renormalization
of the type experienced by massless fermions \cite{MW2}. Weinberg's
result for the power spectrum allows for $\ln(a)$ corrections \cite{SW},
and they do arise from individual terms in Table~\ref{FinTot}. However,
there is no net correction of this form at one loop.

In the end, our null result may not be as surprising as it seems.
It is very simple to show that the scalar self-mass-squared vanishes
at one loop order on a flat background \cite{KW2}. It is also wrong to
think of the scalar as weaker than the fermion; what is weaker is the
fractional one loop correction. Although massless fermions experience
a time-dependent field strength renormalization, this represents a
$\ln(a)$ enhancement of tree order mode functions which fall like
$1/a^{\frac32}$. By contrast, the tree order scalar mode functions
approach a nonzero constant. Indeed, it is the fact that derivatives
of this constant vanish which makes the fractional correction of the
$\kappa h \partial \varphi \partial \varphi$ coupling so small. The sea
of infrared gravitons is present, but it can only couple, at this order,
to the scalar's stress-energy, and the stress-energy of a single scalar
redshifts to zero at late times.

\vskip .5cm

\centerline{\bf Acknowledgements}

We are grateful to A. Linde for illuminating correspondence.
This work was partially supported by NSF grant PHY-0653085 and by the
Institute for Fundamental Theory at the University of Florida.

\end{document}